\documentclass[english,aps,prl,twocolumn,groupedaddress,nofootinbib]{revtex4-1}
\usepackage[LGR,T1]{fontenc}
\usepackage[latin9]{inputenc}
\usepackage{textcomp}
\usepackage{amssymb}
\usepackage{graphicx}
\usepackage{subscript}
\usepackage{amsmath}
\usepackage{color}
\usepackage{ulem}
\makeatletter

\DeclareFontEncoding{LGR}{}{}
\DeclareTextSymbol{\~}{LGR}{126}
\makeatother
\usepackage{babel}

\begin{document}
\title{Unusual interplay between superconductivity and field-induced charge order in YBa$_2$Cu$_3$O$_{y}$}

\author{J. Ka\v{c}mar\v{c}\'{i}k$^{1,2}$,  I.Vinograd$^3$, B.Michon$^{1,4}$, A.Rydh$^5$, A.Demuer$^3$, R.Zhou$^3$, H.Mayaffre$^{3}$, R.Liang$^{6,7}$, W.Hardy$^{6,7}$, D.A.Bonn$^{6,7}$, N.Doiron-Leyraud$^{4}$, L.Taillefer$^{4,7}$, M.-H. Julien$^3$, C.Marcenat$^{8}$ and T.Klein$^{1}$}

\date{\today}
\address{$^1$ Univ. Grenoble Alpes, CNRS, Grenoble INP, Institut N\'eel, F-38000 Grenoble, France}
\address{$^2$ Institute of Experimental Physics, Slovak Academy of Sciences, SK-04001 Ko\v{s}ice, Slovakia}
\address{$^3$ Univ. Grenoble Alpes, INSA Toulouse, Univ. Toulouse Paul Sabatier, CNRS, LNCMI, F-38000 Grenoble, France}
\address{$^4$ Institut quantique,  D\'epartement de physique \& RQMP, Universit\'e de Sherbrooke, Sherbrooke, Qu\'ebec J1K 2R1, Canada}
\address{$^5$ D\'epartment of Physics, Stockholm University, AlbaNova University Center, SE - 106 91 Stockholm, Sweden}
\address{$^6$ Department of Physics and Astronomy, University of British Columbia, Vancouver, BC V6T 1Z1, Canada}
\address{$^7$ Canadian Institute for Advanced Research, Toronto, Ontario M5G 1Z8, Canada}
\address{$^8$ Univ. Grenoble Alpes, CEA, INAC, PhELIQS, LATEQS, F-38000 Grenoble, France \\
Correspondence should be addressed to C.M. (christophe.marcenat@cea.fr), to M.-H.J. (marc-henri.julien@lncmi.cnrs.fr) or to T.K. (thierry.klein@neel.cnrs.fr)}

\date{\today}
\begin{abstract}
We present a detailed study of the temperature ($T$) and magnetic field ($H$) dependence of the electronic density of states (DOS) at the Fermi level, as deduced from specific heat and Knight shift measurements in underdoped YBa$_2$Cu$_3$O$_y$.  We find that the DOS becomes field-independent above a characteristic field $H_{\rm DOS}$ and that the $H_{\rm DOS}(T)$ line displays an unusual inflection near the onset of the long range 3D charge-density wave order. The unusual S-shape of $H_{\rm DOS}(T)$ is suggestive of two mutually-exclusive orders that eventually establish a form of cooperation in order to coexist at low $T$. On theoretical grounds, such a collaboration could result from the stabilisation of a pair-density wave state, which calls for further investigations in this region of the phase diagram.\\
\end{abstract}

\maketitle

There is now compelling evidence that high-$T_c$ superconductivity in cuprates competes with a charge-density wave (CDW) over a substantial range of carrier concentrations~\cite{Wu1,Wu2,Wu3,Chang1,Leboeuf,Ghiringhelli,Comin}. This CDW is most prominent in underdoped YBa$_2$Cu$_3$O$_y$ where it causes a major reconstruction of the Fermi surface \cite{Taillefer,Sebastian2015,DL2007,NDL2007,Sebastian,Ramshaw}. In zero/low magnetic fields, the CDW order is two-dimensional (2D) and short-ranged but it becomes both long-ranged and correlated in all three dimensions (3D) in high magnetic fields~\cite{Wu1,Wu2,Leboeuf,Gerber,Chang2,Jang}. Several consequences of the competition between the CDW and superconducting orders have already been highlighted, including strong field ($H$) and temperature ($T$) dependence of the CDW in the superconducting state~\cite{Wu1,Wu2,Chang1,Ghiringhelli,Gerber,Chang2,Jang}, a diminution of $T_c$ \cite{Liang06} and a severe reduction of the upper critical field $H_{\rm c2}(T=0)$~\cite{Grissonnanche1,Zhou}. 

Here, our specific heat and spin susceptibility measurements reveal an unforeseen effect of this competition: upon cooling, the temperature dependence of the field $H_{\rm DOS}(T)$ above which the electronic density of states (DOS) at the Fermi level saturates, displays a clear inflection when the field-induced long-range CDW order develops~\cite{Leboeuf,Gerber,Chang2,Jang}. $H_{\rm DOS}(T)$ then sharply increases below $\sim 10$ K tending towards $H_{c2}(0)$ for $T\rightarrow 0$~\cite{Marcenat,Zhou}. This results in an unusual S-shape of $H_{\rm DOS}(T)$ which is suggestive of two mutually exclusive orders that eventually establish a form of cooperation in order to coexist at low temperature. These results raise the question as to whether the nature of the superconducting state is altered in order to allow for this collaboration, in which case the low temperature phase could correspond to the predicted pair-density wave (PDW) order \cite{Agterberg1,Berg,Lee1,Fradkin,Wang,Edkins,Chakraborty,Himeda,Raczkowski,Agterberg2,Lee2,Montiel}. 

\begin{figure}
\includegraphics[width=7cm]{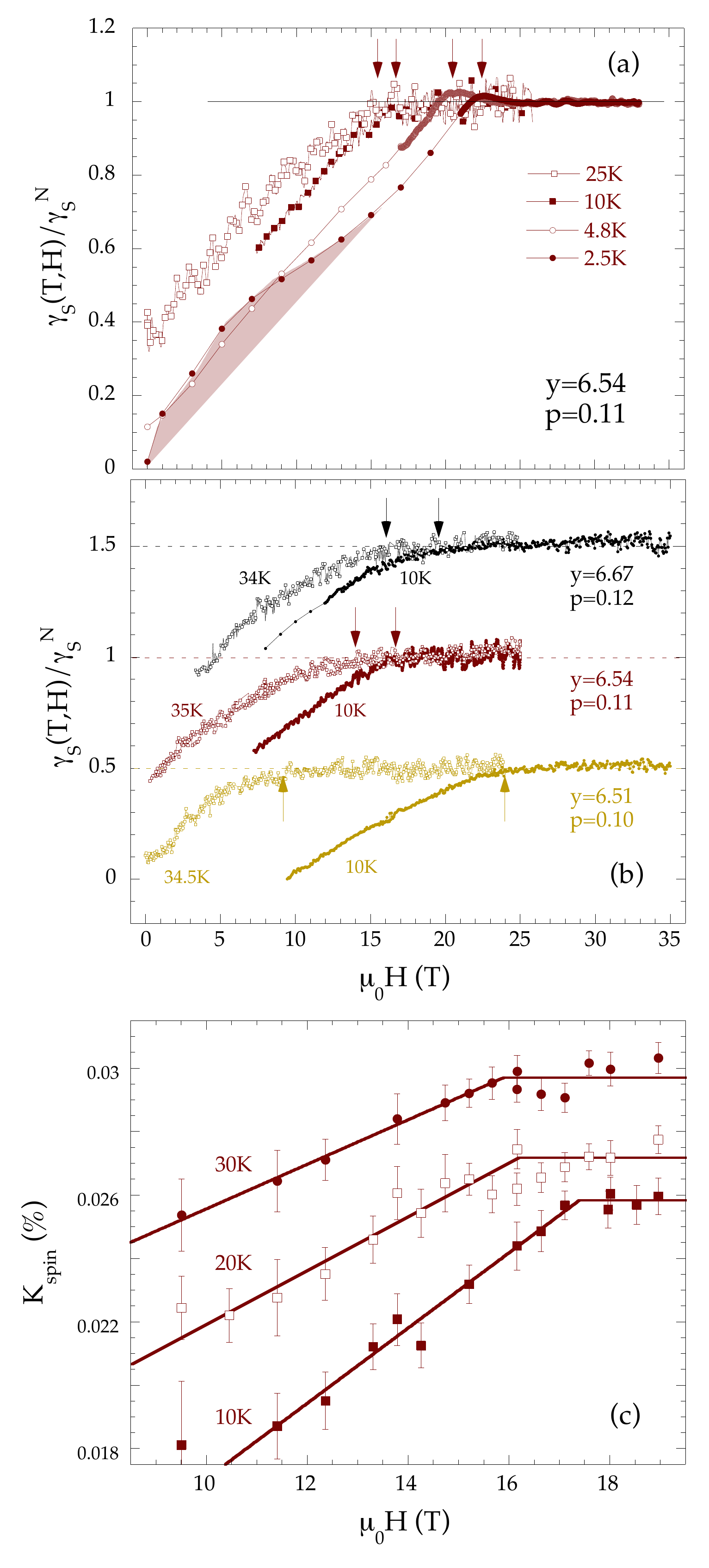}
\caption{(a,b): Magnetic field dependence of the electronic specific heat, $\gamma_S(T,H)$ at different temperatures (the curves in panel b have been shifted for clarity). The arrows represent the field $H_{\rm DOS}(T)$ above which data saturate. At 2.5~K, the data markedly deviate from a linear dependence  (shaded area in panel a, see text for details). (c) Field dependence of the Knight shift for $p=0.109$, solid lines are fits to the data, as explained in the main text.}
\end{figure}

\begin{figure}
\includegraphics[width=7.5cm]{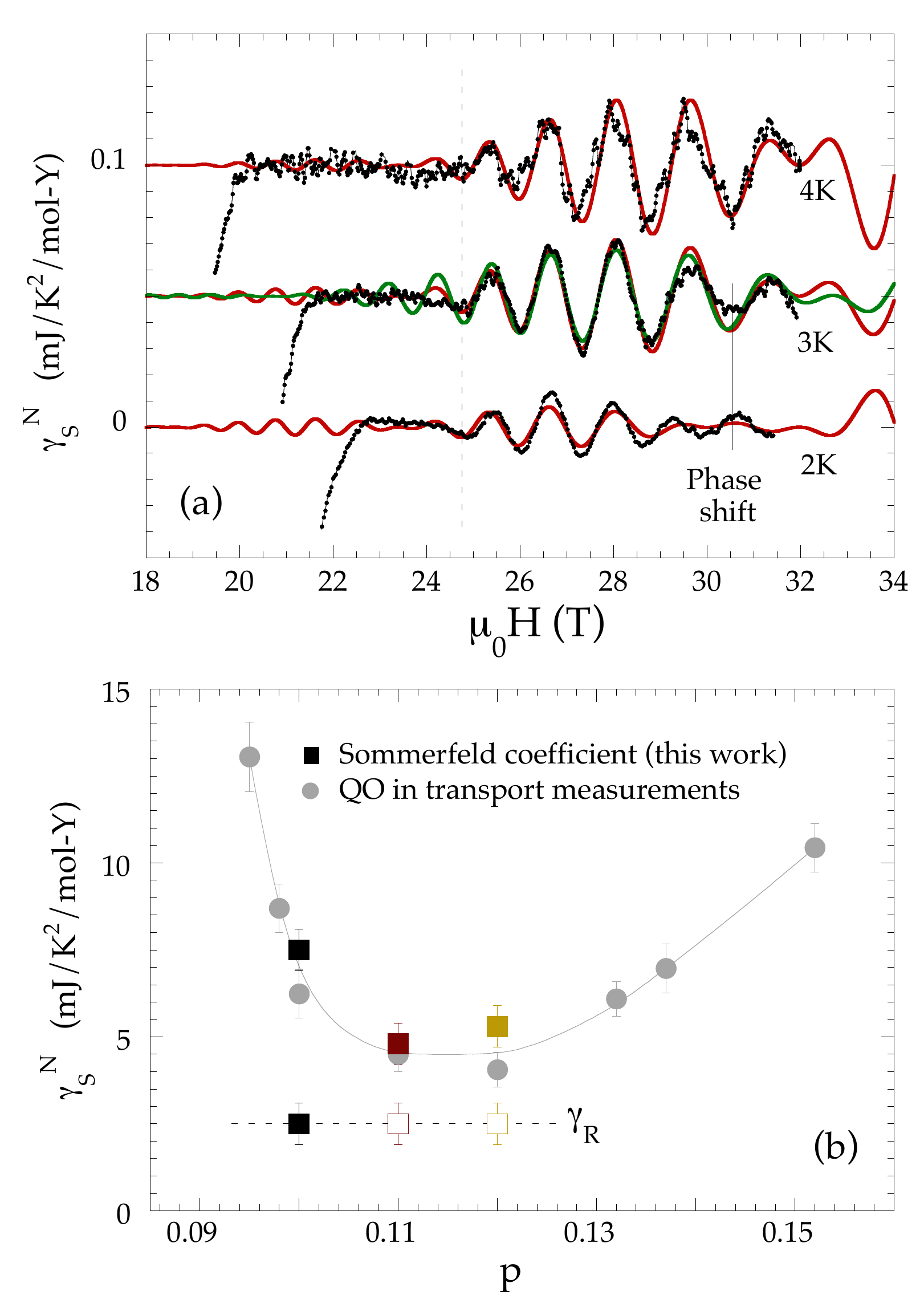}
\caption{(a) Magnetic field dependence of the specific heat at low temperature ($p= 0.11$). Clear quantum oscillations are observed (a smooth polynomial background has been removed from the data and each temperature has been arbitrarily shifted for clarity). Those oscillations can be very well described by the standard Lifshitz-Kosevich formula with $F\sim530$ T and a warping term $t_w \sim15$T (red lines) or $F\sim 520$ T  and $t_w \sim 22$T (green line). (b) $\gamma_S^N$ (closed squares) and residual $C_p/T$ values (open squares) as a function of the doping rate. The doping dependence of the effective mass deduced from quantum oscillations in transport and skin depth measurements \cite{Sebastian,Ramshaw} (grey circles), have also been plotted for comparison (see text for details).}
\end{figure}

 \begin{figure*}
\includegraphics[width=17cm]{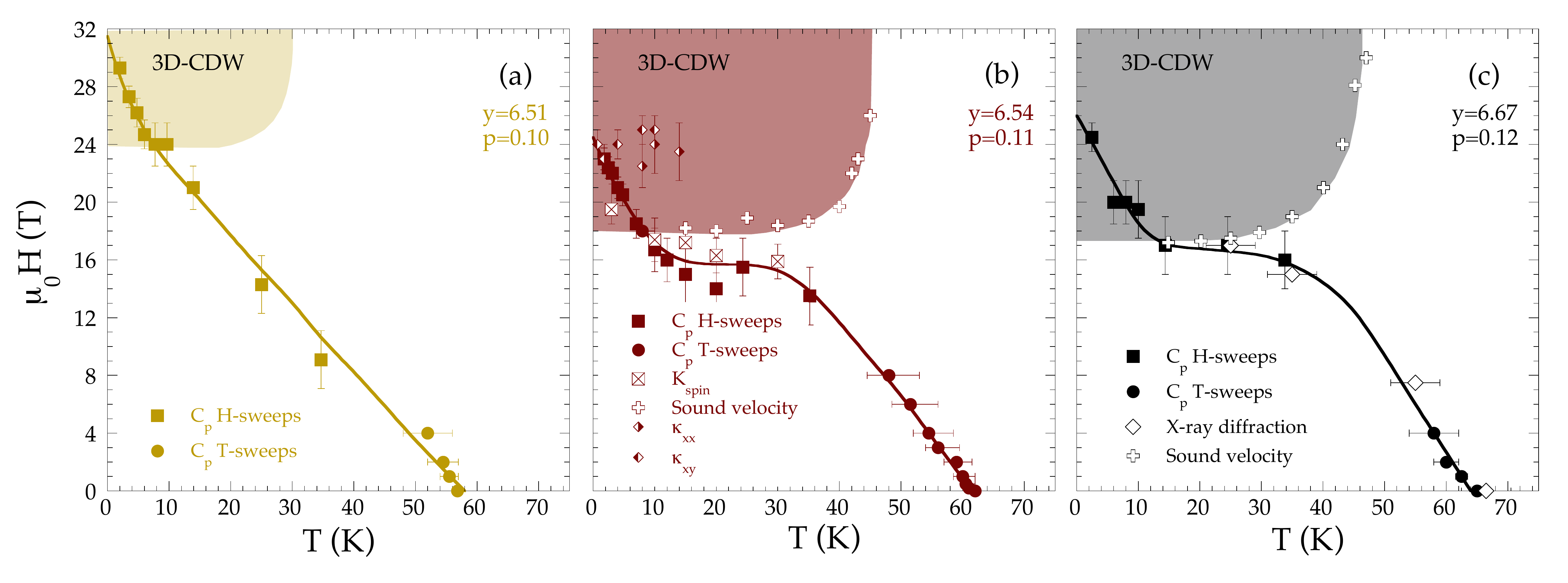}
\caption{$H_{\rm DOS}$ versus temperature for the indicated doping contents. The solid circles and squares have been derived from T (see \cite{Marcenat}) and H sweeps (see Fig.1a-b) of the specific heat, respectively, the crossed squares have been deduced from $K_{\rm spin}(H)$ (see Fig.1c).  As shown a clear "plateau" is observed in $H_{\rm DOS}(T)$ for $p=0.11$ (panel b) and $p=0.12$  (panel c) in the vicinity of the onset of the long range 3D-CDW \cite{Leboeuf2,Grissonnanche2} (open crosses and shaded areas), highlighting the interplay between those two competing orders. For $p = 0.10$ (panel a), a change of slope is observed when $H_{\rm DOS}$ crosses $H_{\rm CDW}$. The field $H_{\rm scat}$ marking the onset of scattering by superconducting fluctuations, as deduced from thermal conductivity measurements (diamonds \cite{Grissonnanche1,Grissonnanche2}) have also been reported for $p=0.11$. Open diamonds in panel (c) correspond to the fields below which the intensity of the CDW diffraction peaks decrease \cite{Chang1}. Lines are guides to the eyes.}
\end{figure*}

\begin{figure}
\includegraphics[width=7.5cm]{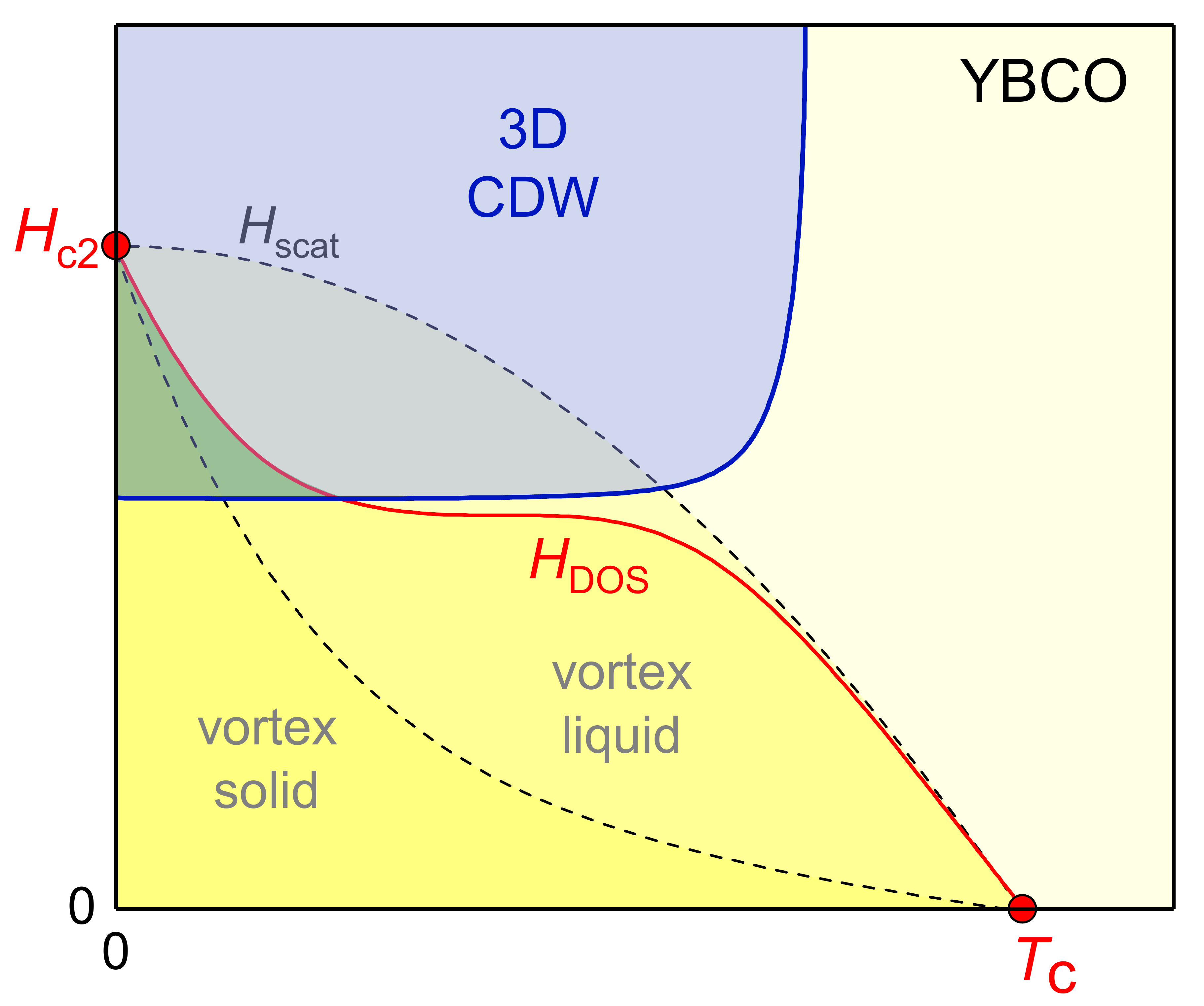}
\caption{Sketch of the $H-T$ diagram of underdoped YBa$_2$Cu$_3$O$_{y}$ emphasazing the interplay between the superconducting and CDW orders. Specific heat and Knight shift measurements show that the density of states at the Fermi level reaches its normal state value above $H_{DOS}$ (see Fig.1). Different shades of yellow tentatively depict the intensity of local superconducting fluctuations, with emphasis on the field scale $H_{scat}$ deduced from thermal conductivity measurements \cite{Grissonnanche1,Grissonnanche2}. The greenish region corresponds to the $H_{CDW}\leq H\leq H_{DOS}$ field range, in which superconductivity might be impacted by the presence of 3D CDW order.}
\end{figure}

The magnetic field dependence of the electronic specific heat of the superconducting phase, $\gamma_S(T,H) = C_p/T-\gamma_R- C_{ph}(T)/T$ has been measured in underdoped YBa$_2$Cu$_3$O$_y$ single crystals \cite{Liang} with doping rates $p=0.10$ ($y=6.51$), $0.11$ ($y=6.54$) and $0.12$ ($y=6.67$)($\gamma_R$ being the residual Sommerfeld coefficient and $C_{ph}(T)$ the phonon contribution, see supplemental materials for technical details). As shown in Fig.1a-b (see also Fig.1 in supplemental materials), $\gamma_S$ becomes field independent above a characteristic field $H_{\rm DOS}(T)$.  For $H\leq H_{DOS}(T)$,  $\gamma_S$ decreases approximatively linearly, except at $2.5$K for which a more complex behaviour is found. As previously reported by Kemper {\it al.} \cite{Kemper}, $\gamma_S$ displays a possible $\sqrt{H}$ dependence at low field and very low temperature (characteristic of $d$-wave superconductors), and crosses over to a more linear dependence above $H\sim 15$T,  but the presence of a Schottky anomaly (which has been subtracted from the data presented in Fig.1a, see technical details in SM) is hindering any detailed discussion of this field dependence. 

Quantum oscillations (QO) are  observed in the $p=0.11$ crystal for $H\geq 25$ T and $T\leq 6$~K (Fig.~2a). The oscillations, here observed down to a field value significantly smaller than in \cite{Riggs}, can be  well described by the Lifshitz-Kosevich (LK) formula (see Eq.~1 in \cite{Riggs}), introducing a frequency $F = 530$~T and a warping term $t_w = 15$~T (as done in \cite{Riggs}, green line in Fig.2a). It is worth noting that the QO abruptly disappear for $H\sim 25$~T (for all  temperatures). As a small change of the parameters ($F = 520$~T, $t_w = 22$~T, red lines in Fig.2a) predicts almost undetectable oscillations below $\sim 25$~T, this dampening of the QO could be due to the presence of a node around 25T. Note however that torque measurements did not indicate the presence of any node in this field range, and suggested that QO can persist well below 25~T \cite{Maharaj}. It is also worth noting that this "onset" field is very close to the field $H_{\rm scat}$ below which the thermal conductivity $\kappa_{xx}$ abruptly decreases~\cite{Grissonnanche1}, marking the onset of strong superconducting fluctuations which are also expected to lead to a significant dampening of the  QO. 

As shown in Fig.~2a (dotted line), a $\pi-$phase shift of the oscillations is observed for $T = 2.3 \pm 0.3$~K and $B\sim 31$~T. Since the LK formula  changes its sign for $k_BT= 0.08\times\hbar qB/m^*$ i.e. for $T/B \sim 0.11\times m_e/m^*$, this phase shift directly implies - without any fitting parameter - that the effective mass is $m^*=1.5\pm0.2$ for $p= 0.11$ (the solid lines in Fig.2a have been calculated with $m^* = 1.5$). In two dimensions, the normal state $\gamma_S^N$ value is directly related to $m^*$ through: $\gamma_S^N \sim 2.9\times(m^*/m_e)$ mJ/molK$^2$/pocket (in two layers systems) and, assuming that the Fermi surface is constituted of one single (electron) pocket per CuO$_2$ layer, one expects $\gamma_S^N \sim 4.4\pm 0.5$ mJ/molK$^2$ in very good agreement with the measured value (see Fig.~2b and Fig.~2 in SM). Note however that we did not take into account the residual specific heat $\gamma_R$ which hence seems not to be directly related to the reconstructed FS (probably related to the CuO chains).  A similar $\gamma_S^N$ (i.e. $m^*$) value is measured for $p=0.12$ but $\gamma_S^N$ increases to $ \sim 7.5\pm 0.5$ mJ/molK$^2$ for $p =0.10$, indicating that the effective mass sharply increases to $2.60\pm0.15$, in good agreement with the change in $m^*$ deduced from quantum oscillations in transport and skin depth measurements \cite{Sebastian,Ramshaw} (grey circles in Fig.2b). 

We now turn to the saturation of the specific heat above $H_{\rm DOS}$, the field for which $\gamma_S$ reaches its maximum value in either temperature or field sweeps (a temperature sweep at 18T - see Fig.2b in SM - confirmed that the two criteria are equivalent). At low temperature, earlier $C_p$~\cite{Marcenat} and NMR Knight shift~\cite{Zhou} measurements showed that the saturation of the DOS coincides with the onset of scattering by superconducting fluctuations (as inferred from thermal conductivity data~\cite{Grissonnanche1}): at $T \simeq$~2-3 K, $H_{\rm DOS}\simeq H_{\rm scat} \equiv H_{c2}(0)\sim 24$ T (see Fig.3a in SM). What do we expect at higher temperatures ? As is well-known in high-$T_c$ cuprates, thermal fluctuations cause the vortex lattice to melt into a vortex liquid at temperatures well below the mean-field transition field $H^{\rm MF}_{\rm c2}$.  As the line tension of vortices vanishes at the melting transition \cite{Fisher,Sudbo}, the normal and vortex-liquid states are the same phase and are connected by a smooth crossover \cite{Fisher,Sudbo}. This means that there is no sharply defined transition at $H^{\rm MF}_{\rm c2}$ for finite $T$ and $H_{DOS}$ can hence not be identified with $H_{c2}$. Nevertheless, the specific heat is still expected to present a smeared anomaly in the vicinity of the former $H^{\rm MF}_{\rm c2}$ line  where most of the ordering energy comes out and the DOS reaches its normal state value. We indeed find that a clear saturation of $C_p(H)$ persists upon heating (Fig.1a-b) but we observe that $H_{\rm DOS}$ rapidly decreases with temperature (Fig.2 and Fig.3). 

Note that a clear "overshoot" is observed in the field dependence of $\gamma_S$ at low temperature (see Fig.1a and Fig.3a in SM). This "overshoot" is reminiscent of the mean field specific heat jump at $H_{c2}^{MF}$. Indeed in the absence of thermal fluctuations, a specific heat jump $\Delta C_p/T =-\mu_0(\partial M/\partial T)_H(dH_{c2}^{MF}/dT)$ is observed at $H=H_{C2}^{MF}$ ($M$ being the reversible magnetisation in the superconducting state) and a smeared "overshoot" is still expected to be observed in presence of thermal fluctuations if $dH_{DOS}/dT \neq 0$ (and the slope of the magnetisation is changing rapidly close to $H_{DOS}$). The observation of such an overshoot at $T/T_c \sim 1/30$ is very unusual as $dH_{c2}^{MF}/dT \rightarrow 0$ at low temperature, but is here thermodynamically consistent with the strong temperature dependence of $H_{DOS}$ below 10K.  

Upon further heating, we find that this decrease of $H_{\rm DOS}(T)$ is followed by a plateau with $H_{\rm DOS} \simeq 15$~T in the 10 - 30~K range, that is, in the vicinity of the onset field of long-range CDW order for both $p=0.11$ and $p = 0.12$ (Fig.3b-c), as detected by sound velocity \cite{Leboeuf,Leboeuf2}, X-ray scattering experiments \cite{Gerber,Chang2,Jang} and thermal Hall conductivity $\kappa_{xy}$ measurements \cite{Grissonnanche2}. Finally above $\sim 30$K, $H_{DOS}$  further decreases (tending towards zero for $T\rightarrow T_c$ and an overshoot is again observed \cite{Marcenat}) giving rise to an unexpected S-shape of the $H_{DOS}(T)$ line.  Note that for $p=0.12$ the $H_{DOS}(T)$ line well agrees with the line below which the intensity of the (short range) CDW diffraction peaks becomes field dependent (open diamonds in Fig.3c), marking the onset of the superconducting phase \cite{Chang1}. For $p=0.10$, in which the CDW onset field is much larger~\cite{Leboeuf2} and the CDW presumably weaker, we do not observe such a plateau. Nevertheless, a change of slope of $H_{\rm DOS}(T)$ remains visible (see Fig.3a) on entering into the CDW phase. 

In order to confirm these results, we have also measured the spin part of the Knight shift $K_{\rm spin}$ in a similarly doped $p= 0.109$ single crystal~\cite{Wu2,Zhou,Zhou2}.  $K_{\rm spin}$ is proportional to the uniform spin susceptibility $\chi_{\rm spin} = \chi_{\rm spin}(q=0,\omega=0)$ at planar sites: $K_{\rm spin} = K_{\rm total} - K_{\rm orb}  = A/g\mu_B\chi_{\rm spin}$, where $ K_{\rm total}$ is the experimentally-measured total Knight shift, $K_{\rm orb}$ the orbital shift, $A$  the hyperfine coupling constant, $g$ the Land\'e factor and $\mu_B$ is Bohr magneton. As discussed in \cite{Zhou}, the field dependence of $\chi_{\rm spin}$ in the superconducting state is expected to reflect changes in the density of states at the Fermi level, even though $\chi_{\rm spin}$ may not be related to the DOS in a simple way. As shown in Figs.1c and Fig.3b, Knight shift measurements unambiguously confirm a saturation of the DOS above $H_{DOS}$. The $K_{\rm spin}(H)$ data have been fitted by a linear increase up to $H_{\rm DOS}$ and to a constant value beyond, with $H_{\rm DOS}$ being itself a fitting parameter and, as shown in Fig.3 (see also Fig.S2 in supplemental materials),. The fits lead to $H_{\rm DOS}$ values in agreement with those obtained in $C_p$ measurements. In particular, the NMR data confirm the nearly constant $H_{\rm DOS}\simeq 16-17$~T for 10~K $\lesssim T \lesssim$~30~K. Note that $K_{spin}$ is not expected to present any overshoot, clearly indicating that the S-shape of the $H_{DOS}$ line is not related to the presence of such an overshoot in $C_p$. Even though the maximum accessible field was limited to 20~T in this NMR experiment, it is important to stress that the saturation is beyond error bars. Furthermore our data in much higher fields ($H\simeq 28$~T) show identical $K_{\rm spin}$ values (see Fig.4 in SM), which demonstrates that the saturation of $K_{\rm spin}$ has indeed already been reached at fields $\sim$~17T.

In principle, the constant DOS could be due to the entering into a new type of gapless superconducting state but we fail to see any theoretical support for such a scenario. Furthermore, our low $T$ data exclude that the saturation results from an accidental compensation between a decrease of the DOS due to the opening of a CDW gap on the one hand and an increase of the DOS in the still-existing superconducting phase on the other hand. If this was the case, $\gamma_S(H)$ would saturate near $H_{\rm CDW}\simeq 17$~T at 2.5~K for $p=0.11$, while we observe that it saturates close to $H_{c2}(0)\simeq24$~T. Finally, as the vortex melting line lies significantly below $H_{\rm DOS}$ (for $T\neq0$, see \cite{Marcenat} for p=0.11 as an exemple), the saturation of the DOS cannot be related to the melting of the vortex solid. Thus, our measurements indicate that there is an intrinsic saturation of the DOS at $E_F$ for $H \geq H_{\rm DOS}(T)$. Note that superconducting fluctuations persist well above $H_{\rm DOS}$ (for $T\neq 0$, see sketch in Fig.4). The plateau apparently reflects a separation between the field scale deduced from probes sensitive to the DOS and the field scale deduced from probes sensitive to vortex scattering. The two field scale collapse as $T \rightarrow 0$ but a detailed discussion of the onset of those fluctuations is well beyond the focus of the present work. The phase diagrams of Fig.~3b-c leave little doubt that the unusual S-shape of $H_{\rm DOS}(T)$ directly results from the influence of three-dimensional (3D) long-range CDW order on  superconductivity (and not from {\it e.g.} a disorder driven Griffiths superconducting phase~\cite{Griffiths1,Griffiths2}). 

The presence of this plateau suggests that high-field CDW and superconductivity initially repel each other, as if they were mutually exclusive orders that cannot coexist \cite{Meier}. However, the upswing of the $H_{\rm DOS}(T)$ line below $\sim$10~K suggests that superconductivity eventually finds a way to accommodate the presence of 3D long-range CDW order. Our measurements do not offer a direct clue on what microscopically characterises the "collaborative" state between CDW and superconducting orders for $T \leq 10$~K and $H\gtrsim 15$~T. However, because unusual effects are seen in the DOS, it is likely that some characteristics of the superconducting state have changed. In this context, it is interesting to note that recent theoretical works \cite{Agterberg1,Lee1,Wang,Chakraborty}, motivated by experimental scanning-tunneling microscopy of the vortex cores \cite{Edkins}, suggested that high magnetic fields may reveal a pair-density-wave (PDW) state in which spatial variations of the superconducting and CDW order parameters are intertwined (see \cite{Fradkin} for a review). While it has been proposed that the 3D CDW is actually a consequence of a primary PDW order \cite{Lee1}, the above-proposed interpretation of our data fits more naturally with the view that PDW order is stabilized only in coexistence region by the presence of independent 3D CDW order \cite{Wang,Chakraborty}. In any event, it would be extremely interesting to compare our results with predictions for the field-dependence of the DOS from different theoretical models. 

\begin{acknowledgments}
 Work in Grenoble was supported by the French Agence Nationale de la Recherche under reference AF-12-BS04-0012-01 (Superfield), by the Laboratoire d'excellence LANEF (ANR-10-LABX-51-01). Part of this work was performed at the LNCMI, a member of the European Magnetic Field Laboratory (EMFL). L.T. acknowledges support from the Canadian Institute for Advanced Research (CIFAR) and funding from the Natural Sciences and Engineering Research Council of Canada (NSERC; PIN:123817), the Fonds de recherche du Qu\'ebec - Nature et Technologies (FRQNT), the Canada Foundation for Innovation (CFI), and a Canada Research Chair. This research was undertaken thanks in part to funding from the Canada First Research Excellence Fund and the Gordon and Betty Moore Foundation's EPiQS Initiative (Grant GBMF5306).  J.K. was supported by the Slovak Research and Development Agency under Grant No. APVV-16-0372.
 \end{acknowledgments}

\end{document}